\documentclass[a4paper,11pt]{article}
\pdfoutput=1 % if your are submitting a pdflatex (i.e. if you have
\usepackage{jheppub} % for details on the use of the package, please
\title{\boldmath  Enhanced Euclidean supersymmetry, 11D supergravity and $SU(\infty)$ Toda equation }

\author[a]{M. Dunajski,}
\author[b]{J. Gutowski,}
\author[c]{and W. A. Sabra}
\affiliation[a]{ Department of Applied Mathematics and Theoretical Physics \\
University of Cambridge \\
Wilberforce Road, Cambridge CB3 0WA, UK.}
\affiliation[b]{Department of Mathematics \\
University of Surrey \\
Guildford, GU2 7XH, UK}
\affiliation[c]{Centre for Advanced Mathematical Sciences and Physics Department \\
American University of Beirut, Beirut, Lebanon}

\emailAdd{m.dunajski@damtp.cam.ac.uk}
\emailAdd{j.gutowski@surrey.ac.uk}
\emailAdd{ws00@aub.edu.lb}

\abstract{
We show how to lift solutions of Euclidean Einstein--Maxwell equations with non--zero cosmological constant to solutions of eleven--dimensional supergravity theory with non--zero fluxes. This yields a class of 11D metrics given in terms of solutions to $SU(\infty)$ Toda equation. We give one example of a regular solution and analyse its supersymmetry.

We  also analyse the integrability conditions of the Killing spinor
equations of $N=2$ minimal gauged supergravity in four Euclidean dimensions. We obtain necessary conditions for the existence
of additional Killing spinors, corresponding to enhancement of
supersymmetry. 
If the Weyl tensor is anti-self-dual then the supersymmetric
metrics satisfying these conditions are given by separable solutions to the $SU(\infty)$ Toda equation. Otherwise
they are ambi--K\"ahler and are conformally equivalent to K\"ahler metrics of Calabi type or
to product metrics on two Riemann surfaces.}

\arxivnumber{1301.1896}

\begin{document}
\maketitle
\flushbottom

\section{Introduction}

Euclidean solutions to Einstein--Maxwell equations with non--zero cosmological constant
(and their lifts to eleven--dimensional supergravity) have recently appeared
in the context of AdS/CFT correspondence. In \cite{sparks1,sparks2} it was argued
that a class of supersymmetric gauge theories on three--dimensional Berger spheres
posses gravity duals given by Euclidean $N=2$ minimal gauged supergravity solutions. See \cite{lukier} for some earlier results on Euclidean supersymmetry. 

The aim of this paper is two--fold.  Firstly we shall show how to lift the Euclidean Einstein--Maxwell space times with
$\Lambda>0$ to solutions of $D=11$ Lorentzian supergravity with non--vanishing fluxes.
The Fubini--Study metric $g_{{\mathbb{C}}P^2}$ on ${\mathbb{C}}P^2$ with the Maxwell field
given by the K\"ahler form leads to an explicit regular eleven--dimensional solution which is
a non--trivial bundle over a product manifold ${\mathbb{C}}P^2\times\widetilde{{\mathbb{C}}P}^3$, where
$\widetilde{{\mathbb{C}}P}^3$ is the non--compact dual to ${\mathbb{C}}P^3$ with the 
Bergmann metric.  
The metric and the four--form are given by
\begin{eqnarray}
ds^2&=&g_{{\mathbb{C}}P^2}+ d\tau^2+e^\tau g_{{\mathbb{R}}^4}-2e^{\tau}(dt+A)(d\chi-\alpha+\frac{1}{2}e^{-\tau}(dt+A)) \nonumber \\
G&=&3\mbox{vol}_{{\mathbb{C}}P^2}- J\wedge F,
\end{eqnarray}
where $g_{{\mathbb{R}}^4}$ is the flat metric on ${\mathbb{R}}^4$ with the K\"ahler form $d\alpha$, the Maxwell field in four dimensions is the K\"ahler form on 
${\mathbb{C}}P^2$ given by $F=dA$, and $J=-d(e^{\tau}(d\chi-\alpha))$. This solution admits a null 
non-hyper-surface-orthogonal isometry ${\partial \over \partial \chi}$.
Our procedure is a modification of the ansatz made by Pope \cite{Pope_lift}, adapted to the Euclidean signature, and anti--self--dual Maxwell fields.  Moreover we show that SUSY solutions in four dimensions in general lift
to non SUSY solutions in eleven dimensions.

Secondly we shall investigate solutions
of $N=2$ minimal gauged supergravity in four Euclidean dimensions (these are the same
as Euclidean Einstein--Maxwell space--times with non--vanishing cosmological constant),
where the Killing spinor equations admit more than one solution. 
In particular, we derive necessary conditions for enhanced 
supersymmetry by analysing the integrability conditions of the
Killing spinor equations.
This will select a subclass
of solutions constructed in \cite{DGST11,DGST11_2}  (see \cite{DH07} and
\cite{GT10} for discussion of supersymmetric solutions to Euclidean $N=2$ SUGRA with $\Lambda=0$). 
In particular the underlying metric has to be ambi--K\"ahler in the sense of
\cite{ACG} w.r.t. both self--dual and anti--self--dual parts of the Maxwell
fields. If the self--dual part of the Maxwell field vanishes,
the integrability conditions impose additional constraints on solutions of the  $SU(\infty)$ Toda equation 
\cite{BF}
\begin{eqnarray}
\label{todaeq}
U_{XX}+U_{YY}+(e^U)_{ZZ}=0, \quad U=U(X, Y, Z)
\end{eqnarray}
which underlies some solutions of \cite{DGST11}.
Here $U_{X}=\partial_X U$ etc.
Imposing these constraints enables us to integrate the $SU(\infty)$ Toda 
equation completely to find that the corresponding solutions are separable, i. e. 
$U(X, Y, Z)=U_1(Z)+U_2(X, Y)$ and thus fall into the class
studied by Tod \cite{T95}.
This class of solutions includes the non--compact analogue of the Fubini--Study metric, $\widetilde{{\mathbb{C}}P}^2=SU(2, 1)/U(2)$ with cosmological constant $\Lambda<0$. In this case the norm of the Maxwell is constant and the solution of (\ref{todaeq}) 
is
\begin{eqnarray}
\label{solution}
U=\ln{\frac{4Z(Z+4\Lambda)}{(1+X^2+Y^2)^2}}.
\end{eqnarray}
If the Maxwell field is neither self--dual nor anti--self--dual, then the integrability conditions imply that the
solutions are type $D$ Euclidean Einstein--Maxwell metrics. They are either conformal rescalings of a product 
metric on two Riemann surfaces, or conformal rescalings of 
\begin{eqnarray}
{\hat{g}} = y h_{\Sigma} + y Q^{-1} dy^2 + y^{-1} Q (d \psi+ \phi)^2
\end{eqnarray}
where $Q(y)$ is a product of two quadratic polynomials in $y$ and
$d\phi$ is a volume form of a metric $h_\Sigma$ on a 2D surface $\Sigma$.
This form of the metric is a special case of the metric appearing in equation (10) of \cite{ACG}.

\section{${\mathbb{C}}P^2$ and $\widetilde{{\mathbb{C}}P}^2$ as SUSY solutions to 
Euclidean gauged supergravity}
Let $\sigma_j, j=1, 2, 3$ be the left--invariant one forms on the group manifold
$SU(2)$ such that $d\sigma_1=\sigma_2\wedge\sigma_3$, etc. The Fubini--Study metric
on ${\mathbb{C}}P^2$ is (see e.g. \cite{Pope78}, \cite{Dbook})
\begin{eqnarray}
\label{cp2}
g_4=f^2 dr^2+\frac{1}{4}(r^2f({\sigma_1}^2+{\sigma_2}^2)+r^2f^2{\sigma_3}^2),
\end{eqnarray}
where $f=(1+\Lambda r^2)^{-1}$ and $\Lambda>0$. This metric is conformally anti--self--dual
(ASD) and Einstein with the cosmological constant $\Lambda$, i. e. $R_{ab}=6\Lambda g_{ab}$. Taking instead $\Lambda<0$ in (\ref{cp2}) gives
the Bergmann metric on the non--compact manifold
$\widetilde{{\mathbb{C}}P}^2=SU(2, 1)/U(2)$.
The metric (\ref{cp2}) is also K\"ahler, albeit with the opposite orientation:
the ASD K\"ahler form is given by
\begin{eqnarray}
\label{cp_maxwell}
F&=&f^2r\;dr\wedge\sigma_3+\frac{1}{2}r^2f\;\sigma_1\wedge\sigma_2\\
&=&d((1/2)r^2 f\;\sigma_3).\nonumber
\end{eqnarray}
In \cite{DGST11} it was shown that supersymmetric solutions
to $N=2$ minimal gauged Euclidean supergravity in four dimensions, with
anti--self--dual Maxwell field, and such that the Killing spinor generates
a Killing vector (which we call $K={\partial \over \partial \psi}$) are of the form\footnote{In this paper
we revert to the standard sign convention for the cosmological constant. Thus the Fubini--Study metric and the round sphere have $\Lambda>0$. In our previous papers
\cite{DGST11,DGST11_2} these metrics had $\Lambda<0$.}
\begin{eqnarray}
\label{prop1_metric}
g=\frac{1}{Z^2}\Big(V(dZ^2+e^U(dX^2+dY^2))+V^{-1}(d\psi+\phi)^2\Big),
\end{eqnarray}
where $U=U(X, Y, Z)$  is a solution of the $SU(\infty)$ Toda equation (\ref{todaeq}),
the function $V$ is given by $4\Lambda V=ZU_Z-2$, and $\phi$ is a 
one--form such that
\begin{eqnarray}
\label{domega}
d\phi=-V_XdY\wedge dX-V_Y dZ\wedge d X -(Ve^U)_ZdX\wedge dY.
\end{eqnarray}
This is also the most general class of ASD Einstein metrics with $\Lambda\neq 0$ and an isometry \cite{T97v1},
\cite{T97v2}.

We can now read off the solution of the $SU(\infty)$ Toda equation from the metric 
(\ref{cp2}).
This can be done in more than one way, as the isometry group of (\ref{cp2})
is $SU(2, 1)$. Thus we make a choice of a left--invariant  Killing vector $K$
such that $i_K\sigma_3=1$ (this would be given by ${\partial \over \partial \psi}$ in the usual
coordinates on $SU(2)$, but there is no need to introduce the coordinates at this stage).
Comparing (\ref{prop1_metric}) with (\ref{cp2}) we find
$Z=-4/(r^2f)$. We now introduce the coordinates $(X, Y)$ on
the two--sphere ${\mathbb{C}}P^1$ such that
\begin{eqnarray}
{\sigma_1}^2+{\sigma_2}^2=\frac{4(dX^2+dY^2)}{(1+X^2+Y^2)^2},
\end{eqnarray}
which yields the expression (\ref{solution}) for $U$ 
and $V=-(Z+4\Lambda)^{-1}$. In these coordinates the ASD K\"ahler
form is
\begin{eqnarray}
\label{f_cp2}
F=\frac{2}{Z^2}dZ\wedge(d\psi+\phi)-\frac{8}{Z}
\frac{dX\wedge dY}{(1+X^2+Y^2)^2}.
\end{eqnarray}
This coincides, up to a constant overall factor, with the ASD Maxwell 
field constructed
in \cite{DGST11,DGST11_2} for a general solution to $SU(\infty)$ Toda 
equation. The formula for $F$ was only 
implicit in these papers  - the explicit expression 
is given by
\begin{eqnarray}
F&=&\frac{\ell^{-1}}{4\Lambda(ZU_Z-2)}(U_{ZY}dZ\wedge dX-U_{ZX}dZ\wedge dY)
\nonumber \\
&+&
\frac{\ell^{-1}}{(ZU_Z-2)^2}(U_{ZX}dX+U_{ZY}dY)\wedge(d\psi+\phi)
\nonumber \\
&+& \frac{\ell^{-1}}{8\Lambda(ZU_Z-2)}e^U(2U_{ZZ}+{U_Z}^2) dX\wedge dY+
\frac{\ell^{-1}}{2(ZU_Z-2)^2}
(2U_{ZZ}+{U_Z}^2) dZ\wedge (d\psi+\phi),
\nonumber \\
\end{eqnarray}
where $\Lambda=-1/(2\ell^{2})$.
Thus we need $\Lambda<0$ for this to be real, and we have established that
$\widetilde{{\mathbb{C}}P}^2$ is super-symmetric.
The Maxwell potential is given by
\begin{eqnarray}
A=\frac{1}{\ell}\Big(\frac{1}{2}\frac{U_Z}{2-ZU_Z}(d\psi+\phi)
+\frac{1}{8\Lambda}(U_X dY-U_Y dX)\Big)
\end{eqnarray}
and the two--form (\ref{f_cp2}) is given by $-4\ell \;dA$.
If instead $\Lambda=1/(2\ell^2)>0$ then (\ref{prop1_metric}) is still supersymmetric
if we replace $F$ by $-F$.

The Maxwell two--form  satisfies $F\wedge F=c. \mbox{vol}$ where $c=\mbox{const}$ 
if
\begin{eqnarray}
e^U({U_Z}^2+2U_{ZZ})^2+4({U_{ZX}}^2+{U_{ZY}}^2)=\frac{1}{4}c e^U
\Big(\frac{ZU_Z-2}{Z}\Big)^4.
\end{eqnarray}
We note that, up to a transformation of $(X, Y)$ coordinates, the function (\ref{solution}) is the most general solution to the above constraint which also satisfies the $SU(\infty)$ Toda equation, and is separable in the sense that $U(X, Y, Z)=U_1(Z)+U_2(X, Y)$.
\vskip5pt
To sum up, if $(M, g)$ is conformally ASD, $F_+=0$ and $F_-$ is covariantly constant then $(M, g)$ is a space of 
constant holomorphic sectional curvature, i.e. a complex space form  
${\mathbb{C}}P^2$, its non--compact dual $\widetilde{{\mathbb{C}}P}^2$, or ${\mathbb{C}}^2$.

\section{Lift to eleven dimensions}
Let $(M_4, g_4, F)$ be a Riemannian  solution of Einstein--Maxwell equations in four dimensions with $\Lambda>0$ and anti--self--dual Maxwell field
$F=dA$, and let $(M_6, g_6, J)$ be a K\"ahler--Einstein  manifold
with the  Ricci tensor  $R_{\alpha\beta}=k_2 {(g_6)}_{\alpha\beta}$ and the 
K\"ahler form $J$.
Let us  consider a Lorentzian metric
\begin{eqnarray}
\label{lift}
ds^2=g_4+g_6-(dt+k_1 A+ B)^2,
\end{eqnarray} 
together with the the four--form 
\begin{eqnarray}
G=\sigma_1 \mbox{vol}_4+ \sigma_2 J\wedge F+ \sigma_3 J\wedge J,
\end{eqnarray}
where $dB=2k_3 J$ and   $k_1, k_2, k_3, \sigma_1, \sigma_2, \sigma_3$ are constants which will be fixed
by the field equations of $D=11$ supergravity
\begin{eqnarray}
\label{einst}
R_{AB}= {1 \over 12} G_{A N_1 N_2 N_3} G_B{}^{N_1 N_2 N_3}
-{1 \over 144} g_{AB} G_{N_1 N_2 N_3 N_4}G^{N_1 N_2 N_3 N_4}
\end{eqnarray}
and
\begin{eqnarray}
\label{gauge}
d \star_{11} G -{1 \over 2} G \wedge G =0.
\end{eqnarray}
Here $R_{AB}$ is the Ricci tensor of $ds^2$, and the capital letter
indices run from $0$ to $10$. The case $\sigma_3=0$ is a modification  
of the ansatz \cite{Pope_lift} adapted to the Euclidean signature. 
We chose the eleven--dimensional volume to be
\begin{eqnarray}
\mbox{vol}_{11}=\frac{1}{6}(dt+k_1A+B)\wedge\mbox{vol}_4\wedge J\wedge J\wedge J
\end{eqnarray}
which yields
\begin{eqnarray}
*_{11} G= (dt+k_1A+B) \wedge \Big({1 \over 6} \sigma_1 J \wedge J \wedge J
-{1 \over 2} \sigma_2 J \wedge J \wedge {{F}} +2 \sigma_3 J \wedge \mbox{vol}_4 \Big).
\end{eqnarray}
We now substitute this to the gauge field equations (\ref{gauge}) and get
the following three conditions for the constants
\begin{eqnarray}
\label{c1}
-k_3 \sigma_2 + {1 \over 6} k_1 \sigma_1 = \sigma_2 \sigma_3, \qquad
\sigma_2 (k_1 + \sigma_2)=0, \qquad \sigma_3( 4 k_3 - \sigma_1)=0.
\end{eqnarray}
The analysis of the Einstein equations is more complicated, and requires the 
computation of the spin connection coefficients. We skip the tedious calculations and only give the answer. Equations (\ref{einst}) hold if and only if
\begin{eqnarray}
\label{c2}
6 k_3^2 = 2 \sigma_3^2+{1 \over 6} \sigma_1^2, \quad k_1^2= \sigma_2^2,
\quad k_2 + 2 k_3^2 = 2 \sigma_3^2 - {1 \over 6} \sigma_1^2, 
\quad 6 \Lambda ={1 \over 3} \sigma_1^2-2 \sigma_3^2.
\end{eqnarray}
Consider the conditions ({\ref{c1}}) together with ({\ref{c2}}). It is straightforward to show that there
are no solutions to these equations with both $\sigma_2 \neq 0$ and $\sigma_3 \neq 0$. 
There are three distinct classes of solutions:

\begin{enumerate}

\item Solutions with $\sigma_2=\sigma_3=0$. These have $k_1=0$. Furthermore one must have
$\Lambda \geq 0$, with
\begin{eqnarray}
\sigma_1^2 = 18 \Lambda, \qquad k_3^2 = {1 \over 2} \Lambda, \qquad k_2 = -4 \Lambda.
\end{eqnarray}
Thus the Maxwell field decouples in this lift, and we end up with an analogue of the Freund--Rubin solution.

\item Solutions with $\sigma_2=0$, $\sigma_3 \neq 0$. These solutions also have $k_1=0$,
and $\Lambda >0$ with
\begin{eqnarray}
k_3^2 = 2 \Lambda, \qquad \sigma_3^2 = -{10 \over 3} \Lambda, \qquad \sigma_1 = \pm 4 \sqrt{2 \Lambda},
\qquad k_2 = -{8 \over 3} \Lambda.
\end{eqnarray}
The Maxwell field also decouples in this case. This is an analogue of the Englert solution.

\item Solutions with $\sigma_2 \neq 0$, $\sigma_3=0$. These solutions have $\Lambda \geq 0$,
with
\begin{eqnarray}
\label{class3}
\sigma_1 = \pm 3 \sqrt{2} \sqrt{\Lambda}, \qquad k_3 = \mp {1 \over \sqrt{2}} 
\sqrt{\Lambda},
\qquad k_2 =-4 \Lambda,
\end{eqnarray}
and $k_1$ and $\sigma_2$ satisfy $k_1 = - \sigma_2$ but are otherwise unconstrained.
This is the most interesting class from our perspective, as the four--dimensional Maxwell
field contributes non--trivially to the flux in eleven dimensions.
\end{enumerate}   
Note that in all cases, only solutions with $\Lambda \geq 0$ can be uplifted.
This should be contrasted with the original Pope ansatz \cite{Pope_lift} where the
Lorentzian Einstein--Maxwell space times with $\Lambda<0$ have been uplifted to eleven dimensions.
\subsection{Example}

We shall consider the third possibility (\ref{class3}), and chose $\sigma_2=-1$
so that the flux is given by
\begin{eqnarray}
\label{flux_lift}
G=3\sqrt{2\Lambda} \mbox{vol}_4- J\wedge F.
\end{eqnarray}
Let us take $M_4={\mathbb{C}}P^2, g_4=g_{{\mathbb{C}}P^2}$ with the ASD Fubini study metric (\ref{cp2}) and the  Maxwell field given by the ASD K\"ahler form (\ref{cp_maxwell}). The six--dimensional K\"ahler--Einstein manifold
is taken to be the non--compact version of the complex projective three space,
$\widetilde{{\mathbb{C}}P}^3$ known as the Bergmann space. The Ricci scalars of ${\mathbb{C}}P^2$ and
$\widetilde{{\mathbb{C}}P}^3$ have the same magnitude but opposite signs.
We shall chose $\Lambda=1/2$, so that the Ricci scalar of ${\mathbb{C}}P^2$ is $12$.

To construct the metric on $\widetilde{{\mathbb{C}}P}^3$ explicitly, consider the metric 
\begin{eqnarray}
|dZ^1|^2+|dZ^2|^2+|dZ^3|^2 -|dZ^4|^2
\end{eqnarray}
on ${\mathbb{C}}^4$. This is $SU(3, 1)$ invariant.
Restricting this metric to the quadric
$|Z^1|^2+|Z^2|^2+|Z^3|^2-|Z^4|^2=-1$
reduces  it to the constant curvature metric on $AdS_7$. 
Both the initial metric, and the quadric constraint are invariant
under $Z^\alpha\rightarrow \exp{(i\theta)}Z^{\alpha}$, and
the Bergmann manifold is the space of orbits under this circle action.
Thus we can express the $AdS_7$ as a non-trivial $U(1)$ 
or ${\mathbb{R}}^*$ bundle over  $\widetilde{{\mathbb{C}}P}^3$:
\begin{eqnarray}
g_{AdS_7}=-(dt+B)^2+g_{\widetilde{{\mathbb{C}}P}^3}, \quad \mbox{where}\quad
J=-dB.
\end{eqnarray}
Finally, the lift (\ref{lift}) of the Fubini--Study metric on 
${\mathbb{C}}P^2$ to the
solution of $D=11$ supergravity is given by a regular metric
\begin{eqnarray}
\label{11_d_metric}
ds^2=g_{{\mathbb{C}}P^2}+g_{\widetilde{{\mathbb{C}}P}^3}-(dt+ A+B)^2,
\end{eqnarray}
which is a non--trivial line bundle over the ten dimensional Riemannian
manifold ${\mathbb{C}}P^2\times \widetilde{{\mathbb{C}}P}^3$ with its product metric. 
This could, if desired, be reduced along the time--like direction
to a $D=10$ Euclidean supergravity.
 Alternatively we can exhibit a  reduction along
a space--like Killing vector in $\widetilde{{\mathbb{C}}P}^3$
which leads to a Lorentzian solution to the type IIA string theory.

It worth remarking that (\ref{11_d_metric}) belongs to the class
of 11D SUGRA solutions with null isometry. To exhibit this isometry,
use local coordinates $(\tau, \chi, p, q, r, s)$ on
$\widetilde{{\mathbb{C}}P}^3$ found in \cite{Pope3} so that the Bergmann metric takes
the form
\begin{eqnarray}
g_{\widetilde{{\mathbb{C}}P}^3}=d\tau^2+e^\tau(dp^2+dq^2+dr^2+ds^2)
+e^{2\tau}(d\chi-pdr-qds)^2,
\end{eqnarray}
In these coordinates $B=e^{\tau}(d\chi-pdr-qds)$, and 
(\ref{11_d_metric}) can be written as
\begin{eqnarray}
\label{final_ds2}
ds^2=g_{{\mathbb{C}}P^2}+ d\tau^2+e^\tau g_{{\mathbb{R}}^4}-2e^{\tau}(dt+A)(d\chi-\alpha+\frac{1}{2}e^{-\tau}(dt+A))
\end{eqnarray}
where $g_{{\mathbb{R}}^4}$ is the flat metric on ${\mathbb{R}}^4$ with the K\"ahler form $d\alpha$. The null isometry is generated by
${\partial \over \partial \chi}$. The corresponding one--form $g({\partial \over \partial \chi}, . )$ is not hyper--surface--orthogonal, and so (\ref{11_d_metric}) is not a plane
wave solution.

\subsection{Supersymmetry of Uplifted ${\mathbb{C}}P^2$ Solution}

The Killing spinor equations of $D=11$ supergravity are
\begin{eqnarray}
\label{kse11}
{\cal{D}}_M \epsilon =0
\end{eqnarray}
where
\begin{eqnarray}
{\cal{D}}_M = \nabla_M -{1 \over 288}\bigg(\Gamma_M{}^{N_1 N_2 N_3 N_4}
G_{N_1 N_2 N_3 N_4}-8 G_{M N_1 N_2 N_3} \Gamma^{N_1 N_2 N_3}\bigg).
\end{eqnarray}
Here $\nabla_M$ is the Levi--Civita connection of the 11D metric $ds^2$,
$\Gamma^M$ are the Dirac matrices in eleven dimensions and $\Gamma^{MNP}=\Gamma^{[M}\Gamma^N\Gamma^{P]}$ etc.
It follows from the work of \cite{sparks3} that the uplifted ${\mathbb{C}}P^2$ solution
is a spin manifold. However, this does not imply that there exists
a spinor satisfying ({\ref{kse11}}).
It is straightforward to show that the uplifted ${\mathbb{C}}P^2$ solution exhibits both a timelike and
null isometry, corresponding to ${\partial \over \partial t}$
and ${\partial \over \partial \chi}$ respectively.
One might therefore attempt to match the uplifted geometry to
the conditions derived in \cite{d11class1} or \cite{d11class2}, in which the conditions
necessary for a solution of D=11 supergravity to preserve the minimal amount of supersymmetry
were derived, by comparing the isometry of the uplifted 
solution to the isometry generated by the vector field dual to the
1-form Killing spinor bilinear associated with any supersymmetric solution. 
This naive matching fails, but this does not constitute a proof
that the solution is not supersymmetric, as the apparent discrepancy in the geometric conditions
may simply be an artefact of a poor choice of gauge.

In order to determine if the uplifted solution is actually supersymmetric, a necessary condition
is that the integrability conditions of the D=11 Killing spinor equations ({\ref{kse11}})
should admit a non-zero solution $\epsilon$. In particular, consider the integrability
conditions
\begin{eqnarray}
\label{supercurv}
{\cal{R}}_{MN} \epsilon = 0,
\end{eqnarray}
where
\begin{eqnarray}
{\cal{R}}_{MN}= [{\cal{D}}_M, {\cal{D}}_N] \ .
\end{eqnarray}
The actual form of the supercovariant curvature ${\cal{R}}_{MN}$ is set out in
\cite{d11integ}. Due to the rather complicated structure of
these integrability conditions, the analysis of ({\ref{supercurv}}) has
been performed using a computer. First, it is straightforward to check that
\begin{eqnarray}
\Gamma^M {\cal{R}}_{MN} \epsilon=0
\end{eqnarray}
for {\it any} spinor $\epsilon$; this also follows as a consequence of the bosonic 
field equations \cite{systematics}.
We choose a basis ${\bf{e}}^0, {\bf{e}}^\mu, {\bf{e}}^a$,
where
\begin{eqnarray}
{\bf{e}}^0 = dt +A+B
\end{eqnarray}
and let ${\bf{e}}^\mu$ be an appropriately chosen basis for ${\mathbb{C}}P^2$, and
${\bf{e}}^a$ be a basis for $\widetilde{{\mathbb{C}}P}^3$.
Using some further computer analysis, one finds that the integrability condition
({\ref{supercurv}}) is equivalent to
\begin{eqnarray}
\label{supercurv2}
F_{\mu \nu}\Gamma^{\mu \nu} \epsilon =0, \quad \Gamma_0 J_{ab} \Gamma^{ab} \epsilon = 6 \epsilon.
\end{eqnarray}
These conditions reduce the number of real degrees of freedom in $\epsilon$
from $32$ down to $4$, so the solution can preserve at most $N=4$ supersymmetry.
We remark that these conditions also imply that
\begin{eqnarray}
G_{N_1 N_2 N_3 N_4} \Gamma^{N_1 N_2 N_3 N_4} \epsilon = -72 \epsilon
\end{eqnarray}
and using this identity the Killing spinor equation ({\ref{kse11}}) simplifies to
\begin{eqnarray}
\label{kse11b}
\bigg(\nabla_M +{1 \over 24}G_{M N_1 N_2 N_3}\Gamma^{N_1 N_2 N_3}
+{1 \over 4} \Gamma_M \bigg) \epsilon =0.
\end{eqnarray}
It remains to analyse ({\ref{kse11b}}), making use of the conditions ({\ref{supercurv2}}).
The non-zero components of the $D=11$ spin connection are
\begin{eqnarray}
\Omega_{0, \mu \nu}=\Omega_{\mu,0\nu} ={1 \over 2} F_{\mu \nu},
\quad
\Omega_{0,ab}=\Omega_{a,0b}=-{1 \over 2}J_{ab} \ ,
\nonumber  \\
\Omega_{\mu, \nu \rho} = {\tilde{\omega}}_{\mu, \nu \rho},
\quad
\Omega_{a,bc} = {\buildrel{\circ} \over {\omega}}_{a,bc},
\end{eqnarray}
where ${\tilde{\omega}}_{\mu, \nu \rho}$ is the spin connection of
${\mathbb{C}}P^2$, and ${\buildrel{\circ} \over {\omega}}_{a,bc}$  is the spin 
connection of $\widetilde{{\mathbb{C}}P}^3$.

Then the $M=0$ component of ({\ref{kse11b}}) implies that
\begin{eqnarray}
\partial_t \epsilon = - \Gamma_0 \epsilon
\end{eqnarray}
so
\begin{eqnarray}
\label{sol1a}
\epsilon = \bigg( \cos t {\bf 1}- \sin t \Gamma_0  \bigg) {\hat{\epsilon}}
\end{eqnarray}
where
$
\partial_t {\hat{\epsilon}}=0 \ .
$
The conditions ({\ref{supercurv2}}) are equivalent to
\begin{eqnarray}
\label{supercurv3}
F_{\mu \nu}\Gamma^{\mu \nu} {\hat{\epsilon}} =0, \quad \Gamma_0 J_{ab} \Gamma^{ab} {\hat{\epsilon}} = 6 {\hat{\epsilon}}.
\end{eqnarray}
Next, consider the $M=\mu$ component of ({\ref{kse11b}}). This is equivalent to
\begin{eqnarray}
\bigg(\partial_\mu +{1 \over 4} \omega_{\mu, \nu_1 \nu_2} \Gamma^{\nu_1 \nu_2}
\bigg) \epsilon -{1 \over 2}\Gamma_\mu \epsilon -F_{\mu \lambda}\Gamma_0 \Gamma^\lambda
\epsilon =0.
\end{eqnarray}
On substituting ({\ref{sol1a}}) into this expression, and evaluating the terms
dependent on $\sin t$ and $\cos t$ independently, one obtains the condition
\begin{eqnarray}
 {1 \over 2}\Gamma_\mu {\hat{\epsilon}} +F_{\mu \lambda}\Gamma_0 \Gamma^\lambda
{\hat{\epsilon}} =0 \ .
\end{eqnarray}
On contracting this expression with $\Gamma^\mu$, and using ({\ref{supercurv3}}), one
finds
$
{\hat{\epsilon}}=0 \ .
$
It follows that the uplifted solution (\ref{final_ds2}, \ref{flux_lift}) is not supersymmetric.

\section{Solutions with Enhanced Supersymmetry}
\label{section5}

In this section, we examine solutions of minimal Euclidean gauged supergravity
with enhanced supersymmetry. We shall consider solutions which admit
a spinor $\epsilon$ satisfying the following Killing spinor equation:
\begin{eqnarray}
\label{kse}
\left( \partial _{\mu }+{\frac{1}{4}}\Omega _{\mu ,\nu _{1}\nu _{2}}\Gamma
^{\nu _{1}\nu _{2}}+{\frac{i}{4}}F_{\nu _{1}\nu _{2}}\Gamma ^{\nu _{1}\nu
_{2}}\Gamma _{\mu }+{\frac{1}{2\ell }}\Gamma _{\mu }-{\frac{i}{\ell }}A_{\mu
}\right) \epsilon =0.  
\end{eqnarray}
In this case, the cosmological constant is given by $\Lambda =-{1 \over 2 \ell^2}$,
and we do not assume that the Maxwell field strength $F=dA$ is either self or anti-self-dual.

All supersymmetric solutions preserving one quarter of the supersymmetry 
were classified in \cite{DGST11_2}, and 
 one finds that
the metric and gauge potential are given by 
\begin{eqnarray}
g &=&2\lambda ^{2}\sigma ^{2}(d\psi +\phi )^{2}+{\frac{1}{\lambda
^{2}\sigma ^{2}}}\left( {\frac{1}{2}}dx^{2}+2 e^{2u} dzd\bar{z}\right)  \nonumber \\
\nonumber \\
A &=&{\frac{1}{\sqrt{2}}}(\lambda ^{2}-\sigma ^{2})(d\psi +\phi )-
{\frac{i\ell }{2}}\partial _{z} u dz+{\frac{i\ell }{2}}\partial _{\bar{z}} u d\bar{z} 
 \label{fmm1}
\end{eqnarray}
where $\lambda ,\sigma ,u$ are functions, and $\phi =\phi _{x}dx+\phi
_{z}dz+\phi _{\bar{z}}d\bar{z}$ is a 1-form. All components of the metric
and gauge potential are independent of the co-ordinate $\psi $; and $u$, $\lambda $, $\sigma 
$, $\phi $ must satisfy
\begin{equation}
\partial_x u =-{\frac{1}{\sqrt{2}\ell }}(\lambda ^{-2}+\sigma ^{-2})
\label{cc1x}
\end{equation}
and 
\begin{eqnarray}
\partial _{z}\partial _{\bar{z}} (\lambda^{-2}-\sigma^{-2})
+ e^{2u} \bigg( \partial_x^2 (\lambda^{-2}-\sigma^{-2})
+3 (\lambda^{-2}-\sigma^{-2}) \partial_x^2 u
\nonumber \\
+3 (\lambda^{-2}-\sigma^{-2}) (\partial_x u)^2 +3 \partial_x u \partial_x (\lambda^{-2}-\sigma^{-2})
+{1 \over 2} \ell^{-2} (\lambda^{-2}-\sigma^{-2})^3 \bigg)=0
\label{cc2x}
\end{eqnarray}
and 
\begin{equation}
\partial _{z}\partial _{\bar{z}} u + e^{2u} \bigg( \partial_x^2 u + {1 \over 2} (\partial_x u)^2
+{3 \over 4} \ell^{-2}(\lambda^{-2}-\sigma^{-2})^2 \bigg)=0
\label{cc3x}
\end{equation}
and 
\begin{eqnarray}
d\phi &=&-{\frac{i}{(\lambda \sigma )^{2}}}(\partial _{z}\log \frac{\lambda 
}{\sigma })dx\wedge dz+{\frac{i}{(\lambda \sigma )^{2}}}(\partial _{\bar{z}}
\log \frac{\lambda }{\sigma })dx\wedge d\bar{z}  \nonumber \\
&&+{\frac{i e^{2u}}{(\lambda \sigma )^{2}}}\left( 2\partial _{x}\log 
\frac{\sigma }{\lambda }+\sqrt{2}\ell ^{-1}\left( \frac{{\lambda ^{2}}-
{\sigma ^{2}}}{(\lambda \sigma )^{2}}\right) \right) dz\wedge d\bar{z}\ .  \label{cc4x}
\end{eqnarray}
We remark that the self-dual and anti-self-dual parts of $F$ are given by:
\begin{eqnarray}
F^- &=& -2 (d \psi + \phi) \wedge \big(\sqrt{2} \lambda d \lambda +{1 \over 2 \ell} dx \big)
-{\sqrt{2}i e^{2u} \over \lambda^2 \sigma^2} \big(2 \lambda \partial_x \lambda+{1 \over \sqrt{2} \ell}
\big) dz \wedge d {\bar{z}}
\nonumber \\
&-&{\sqrt{2} i \over \lambda \sigma^2} dx \wedge \big(\partial_z \lambda dz - \partial_{\bar{z}} \lambda
d {\bar{z}} \big)
\end{eqnarray}
and
\begin{eqnarray}
F^+ &=& 2 (d \psi + \phi) \wedge \big(\sqrt{2} \sigma d \sigma +{1 \over 2 \ell} dx \big)
-{\sqrt{2}i e^{2u} \over \lambda^2 \sigma^2} \big(2 \sigma \partial_x \sigma+{1 \over \sqrt{2} \ell}
\big) dz \wedge d {\bar{z}}
\nonumber \\
&-&{\sqrt{2} i \over \sigma \lambda^2} dx \wedge \big(\partial_z \sigma dz - \partial_{\bar{z}} \sigma
d {\bar{z}} \big) \ .
\end{eqnarray}
On setting $(F^\pm)^2=F^\pm_{\mu \nu} F^{\pm \mu \nu}$, one obtains
\begin{eqnarray}
\label{squares}
(F^+)^2 &=& {4 \over \ell^2} \bigg( 8 \ell^2 e^{-2u} \sigma^2 \partial_z \sigma \partial_{\bar{z}} \sigma
+(1+2 \sqrt{2} \ell \sigma \partial_x \sigma)^2 \bigg)
\nonumber \\
(F^-)^2 &=& {4 \over \ell^2} \bigg( 8 \ell^2 e^{-2u} \lambda^2 \partial_z \lambda \partial_{\bar{z}} \lambda
+(1+2 \sqrt{2} \ell \lambda \partial_x \lambda)^2 \bigg).
\end{eqnarray}
In addition, the conditions ({\ref{cc1x}}), ({\ref{cc2x}}) and ({\ref{cc3x}}) imply
\begin{eqnarray}
\label{lseq}
e^{-2u} \partial_z \partial_{\bar{z}} \sigma^{-2} + \partial_x \partial_x \sigma^{-2} -{3 \sqrt{2} \over \ell}
\sigma^{-2} \partial_x \sigma^{-2} + 2 \ell^{-2} \sigma^{-6} &=&0
\nonumber \\
e^{-2u} \partial_z \partial_{\bar{z}} \lambda^{-2} + \partial_x \partial_x \lambda^{-2} -{3 \sqrt{2} \over \ell}
\lambda^{-2} \partial_x \lambda^{-2} + 2 \ell^{-2} \lambda^{-6} &=&0.
\end{eqnarray}

The integrability conditions of ({\ref{kse}}) can be decomposed into positive and
negative chirality parts
\begin{eqnarray}
\label{intg}
\bigg( {1 \over 4} W^\pm_{\mu \nu \lambda_1 \lambda_2} \Gamma^{\lambda_1 \lambda_2}
+{i \over 2 \ell} F^\pm_{\lambda_1 [\mu} g_{\nu] \lambda_2} \Gamma^{\lambda_1 \lambda_2}
-{i \over \ell} F^\pm_{\mu \nu} \bigg) \epsilon_\mp
+{i \over 2} \nabla_\sigma F^\pm_{\mu \nu} \Gamma^\sigma \epsilon_\pm=0
\end{eqnarray}
where $W^\pm$ are the self-dual and anti-self dual parts of the Weyl tensor $W$, with $W=W^+ + W^-$,
and
\begin{eqnarray}
\gamma_5 \epsilon_\pm = \pm \epsilon_\pm \ .
\end{eqnarray}
We assume that there exists a Killing spinor $\epsilon^1$ satisfying ({\ref{kse}}) and its associated integrability conditions ({\ref{intg}}). The components of $\epsilon^1$ are the functions
$\lambda,  \sigma$ in an appropriately chosen gauge, using spinorial geometry techniques
as described in \cite{DGST11_2}.

To begin, we consider solutions preserving half of the supersymmetry. We denote the
additional spinor by $\epsilon^2$, 
and it is particularly convenient to write the components of $\epsilon^2$ as
\begin{eqnarray}
\epsilon^2_+ &=& \alpha \epsilon^1_+ + \beta C* \epsilon^1_+
\nonumber \\
\epsilon^2_- &=& \theta \epsilon^1_- + \rho C* \epsilon^1_-
\end{eqnarray}
where $\alpha, \beta, \theta, \rho$ are complex functions.

On evaluating ({\ref{intg}})
acting on  $\epsilon^2_\pm$ we eliminate the Weyl tensor terms using
the conditions on $\epsilon^1_\pm$, to obtain
\begin{eqnarray}
\label{pos1}
-{1 \over 2}(\alpha-\theta) \nabla_\sigma F^-_{\mu \nu} \Gamma^\sigma \epsilon^1_-
+{1 \over 2} (\beta+\rho) \nabla_\sigma F^-_{\mu \nu} \Gamma^\sigma C* \epsilon^1_-
\nonumber \\
+ \ell^{-1} \beta \big(  F^-_{\lambda_1 [\mu} g_{\nu] \lambda_2} \Gamma^{\lambda_1 \lambda_2}
-2 F^-_{\mu \nu} \big) C* \epsilon^1_+ =0
\end{eqnarray}
and
\begin{eqnarray}
\label{neg1}
{1 \over 2}(\alpha-\theta) \nabla_\sigma F^+_{\mu \nu} \Gamma^\sigma \epsilon^1_+
+{1 \over 2} (\beta+\rho) \nabla_\sigma F^+_{\mu \nu} \Gamma^\sigma C* \epsilon^1_+
\nonumber \\
+ \ell^{-1} \rho \big(  F^+_{\lambda_1 [\mu} g_{\nu] \lambda_2} \Gamma^{\lambda_1 \lambda_2}
-2 F^+_{\mu \nu} \big) C* \epsilon^1_- =0.
\end{eqnarray}
Note that contracting ({\ref{pos1}})
with $F^{- \mu \nu}$ and contracting ({\ref{neg1}}) with $F^{+ \mu \nu}$ one finds
\begin{eqnarray}
\label{xx1}
-{1 \over 4}(\alpha-\theta) \nabla_\sigma (F^-)^2 \Gamma^\sigma \epsilon^1_-
+{1 \over 4} (\beta+\rho) \nabla_\sigma (F^-)^2 \Gamma^\sigma C* \epsilon^1_-
-2 \ell^{-1} \beta (F^-)^2 C* \epsilon^1_+ =0
\nonumber \\
\end{eqnarray}
and
\begin{eqnarray}
\label{xx2}
{1 \over 4}(\alpha-\theta) \nabla_\sigma (F^+)^2 \Gamma^\sigma \epsilon^1_+
+{1 \over 4} (\beta+\rho) \nabla_\sigma (F^+)^2 \Gamma^\sigma C* \epsilon^1_+
-2 \ell^{-1} \rho (F^+)^2 C* \epsilon^1_- =0
\nonumber \\
\end{eqnarray}
respectively.

There are a number of possible cases. 
First, if both $F^+=0$ and $F^-=0$; then $W^+=0$ and $W^-=0$ as a consequence of ({\ref{intg}}).
The condition $W=0$ together with the requirement that the metric be Einstein imply in this
case that the manifold must be locally isometric to $H^4$.
Henceforth, we shall assume that $F \neq 0$.

Suppose also that $\alpha-\theta=0$ and $\beta+\rho=0$. Then ({\ref{xx1}}) and ({\ref{xx2}}) imply that
either $F=0$ or $\beta=\rho=0$. Discarding the case $F=0$, if $\beta=\rho=0$ and $\theta=\alpha$ then
\begin{eqnarray}
\epsilon^2 = \alpha \epsilon^1.
\end{eqnarray}
However, as both $\epsilon^2$ and $\epsilon^1$ satisfy ({\ref{kse}}), this implies that $\alpha$ must be constant. In this case, there is no supersymmetry enhancement. 

To proceed, suppose that $F^- \neq 0$. Then on using ({\ref{xx1}}) to solve for $C* \epsilon^1_+$ in terms of 
$\epsilon^1_-$ and $C* \epsilon^1_-$, and substituting the resulting expression back into ({\ref{pos1}}), one
can rewrite ({\ref{pos1}}) as
\begin{eqnarray}
\label{auxconf1}
\bigg((F^-)^2 \nabla_\sigma F^-_{\mu \nu}
-{1 \over 2} F^-_{\mu \nu} \nabla_\sigma (F^-)^2  +{1 \over 2} F^-_{\sigma [\mu}
\nabla_{\nu]} (F^-)^2 
-{1 \over 2} g_{\sigma [\mu} {F}^-_{\nu] \lambda} \nabla^\lambda (F^-)^2  \bigg) \Gamma^\sigma
{\hat{\epsilon}} =0
\nonumber \\
\end{eqnarray}
where
\begin{eqnarray}
{\hat{\epsilon}}=(\alpha - \theta) \epsilon^1_- - (\beta+\rho) C* \epsilon^1_- \ .
\end{eqnarray}
The spinor ${\hat{\epsilon}}$ is non-zero, as $\alpha-\theta$ and $\beta+\rho$ cannot both vanish. 
The condition ({\ref{auxconf1}}) then implies that
\begin{eqnarray}
\label{ambi1}
\nabla^- F^-=0
\end{eqnarray}
where $\nabla^-$ is the Levi-Civita connection of the conformally rescaled metric
$g^- = |F^-| g$.
Similarly, if $F^+ \neq 0$, one finds that
\begin{eqnarray}
\label{ambi2}
\nabla^+ F^+ =0
\end{eqnarray}
where $\nabla^+$ is the Levi-Civita connection of $g^+ = |F^+| g$. 
It follows that if both $F^+ \neq 0$ and $F^- \neq 0$ then the manifold is ambi-K\"ahler.
Such geometries have been classified in \cite{ACG}. However, the integrability conditions
of the Killing spinor equations ({\ref{intg}}) impose a number of additional conditions.

In particular, suppose that $F^- \neq 0$, and evaluate the condition $\nabla^- F^-=0$
explicitly, substituting in the expression for $|F^-|$ given in ({\ref{squares}}).
One obtains a set of PDEs for $\lambda$, and after some rather involved
manipulation, one finds that if $\partial_z \lambda \neq 0$ then 
$\partial_x \lambda=0$. However, ({\ref{lseq}}) then implies that 
$\partial_x u=0$, which is inconsistent with ({\ref{cc1x}}).
Similarly, if $F^+ \neq 0$ then one finds that $\partial_z \sigma =0$.
Conversely, $F^-=0$ implies that $\partial_z \lambda =0$, and $F^+=0$ implies
that $\partial_z \sigma=0$ as a consequence of ({\ref{squares}}).

Hence the conditions 
\begin{eqnarray}
\label{chm}
\partial_z \lambda = \partial_z \sigma=0
\end{eqnarray}
are necessary for the enhancement of supersymmetry.
However, they are not sufficient, and an example of a
quarter-supersymmetric solution found in \cite{sparks3}
for which ({\ref{chm}}) holds is constructed in the following
section.

Using ({\ref{chm}}), the equations ({\ref{lseq}}) can then be solved for $\sigma, \lambda$ to give
\begin{eqnarray}
\sigma^{-2} = -{\ell \over \sqrt{2}} \bigg( {k_1 x + k_2 \over {1 \over2} k_1 x^2+k_2 x + k_3} \bigg),
\quad \lambda^{-2} = -{\ell \over \sqrt{2}} \bigg( {n_1 x + n_2 \over {1 \over2} n_1 x^2+n_2 x + n_3} \bigg)
\end{eqnarray}
for constants $k_1, k_2, k_3, n_1, n_2, n_3$. This satisfies ({\ref{cc2x}}).
The condition ({\ref{cc1x}}) implies that
\begin{eqnarray}
\label{u_general}
u = {1 \over 2} \log \bigg( \big( {1 \over2} k_1 x^2+k_2 x + k_3 \big)\big( {1 \over2} n_1 x^2+n_2 x + n_3 \big) \bigg)
+ {\cal{G}}(z, {\bar{z}})
\end{eqnarray}
where ${\cal{G}}$ is a real function of $z, {\bar{z}}$.
The equation ({\ref{cc3x}}) is equivalent to
\begin{eqnarray}
\partial_z \partial_{{\bar{z}}} {\cal{G}} +{1 \over 2} (k_1 n_3 + n_1 k_3 - k_2 n_2) e^{2 {\cal{G}}}=0
\end{eqnarray}
and the metric is
\begin{eqnarray}
\label{metric_14}
g = {4 \over \ell^2} \big((k_1 x + k_2)(n_1 x + n_2) \big)^{-1} W^{-1} (d \psi + \phi)^2
\nonumber \\
+ \big((k_1 x + k_2)(n_1 x + n_2) \big) \bigg( {\ell^2 \over 4} W dx^2 +{1 \over 2} \ell^2 ds_{M_2}^2 \bigg)
\end{eqnarray}
where
\begin{eqnarray}
W = {1 \over ({1 \over2} k_1 x^2+k_2 x + k_3)({1 \over2} n_1 x^2+n_2 x + n_3)}
\end{eqnarray}
and
\begin{eqnarray}
ds_{M_2}^2 = 2 e^{2 {\cal{G}}} dz d {\bar{z}} \ .
\end{eqnarray}
$M_2$ is either $S^2, {\mathbb{R}}^2$ or $H^2$, according as to
whether $k_1 n_3 + n_1 k_3 - k_2 n_2$ is positive, zero or negative respectively.
Also,
\begin{eqnarray}
d \phi = {i \over 2} \ell^2 (n_1 k_2 - k_1 n_2) e^{2 {\cal{G}}} dz \wedge d {\bar{z}} \ .
\end{eqnarray}
We remark that the case for which $M_2=S^2$ has been obtained in \cite{sparks3}.
The relationship between the solutions found here and those in \cite{sparks3}
will be investigated in greater detail in the following section.

\subsubsection*{ASD case and separable solutions to $SU(\infty)$ Toda equation.}
The special cases $F^-=0$ and $F^+=0$ correspond to 
$n_2^2 -2 n_1 n_3=0$ and $k_2^2 -2 k_1 k_3 =0$ respectively.
In the ASD case, the metric $g$ is Einstein, but not K\"ahler, whereas
the metric $g^- = |F^-|g$ is K\"ahler but not Einstein.

If $n_1\neq 0$ then the $x$--dependence in $u$ in  (\ref{u_general}) is given by a logarithm of a quartic
with one repeated real root, and two other roots. Without loss of generality
we may translate $x$ to set this repeated root to $x=0$. In the ASD case we have
\begin{eqnarray}
\partial_z\lambda=0, \quad \partial_x\lambda=-\frac{1}{2\sqrt{2}\ell}\lambda^{-1}
\end{eqnarray}
and (\ref{cc2x}) is implied by (\ref{cc1x}) and (\ref{cc3x}).
Now the transformation
\begin{eqnarray}
x=\frac{1}{Z}, \quad z=\frac{1}{2}(X+iY), \quad u=\frac{U}{2}-2\log{Z}
\end{eqnarray}
reduces (\ref{cc3x}) to the $SU(\infty)$ Toda equation (\ref{todaeq}) for $U(X, Y, Z)$.
The corresponding solution is separable \cite{T95, MSW}
\begin{eqnarray}
U(X, Y, Z)=U_1(Z)+U_2(X, Y), \quad\mbox{where}\quad U_1(Z)=\log{(\alpha Z^2+\beta Z+\gamma)}
\end{eqnarray}
and $U_2(X, Y)$ satisfies the Liouville equation
\begin{eqnarray}
(U_2)_{XX}+(U_2)_{YY}+2\alpha e^{U_2}=0.
\end{eqnarray}
Here $\alpha, \beta, \gamma$ are constants which depend on $(n_1, n_2, n_3)$. 
 The resulting metric is 
\begin{eqnarray}
g=\frac{1}{Z^2}\Big(V (\alpha Z^2+\beta Z+\gamma)^2 h_3+V^{-1}(d\psi +\phi)^2)
\end{eqnarray}
where the three-metric $h_3$ is Einstein so its curvature is constant and
\begin{eqnarray}
V=-\frac{1}{4\Lambda}\Big(\frac{\beta Z+2\gamma}{\alpha Z^2+\beta Z+\gamma}\Big).
\end{eqnarray}
If
$\alpha=0$ then $U_2$ is harmonic and can be set to zero by
a coordinate transformation. The metric $g_3$ is then hyperbolic if $\beta\neq 0$ or flat
if $\beta=0$.
The case $(\alpha\neq0, \gamma=0, \beta/\alpha=4\Lambda)$ corresponds to the $\widetilde{{\mathbb{C}}P}^2$ solution (\ref{solution}). In this case $n_1=0$ and $|F^-|$ is constant.

\subsubsection*{General case and ambi-K\"ahler surfaces of Calabi type.}
More generally, the geometry (\ref{metric_14}) corresponds to a K\"ahler surface of Calabi type,
as described in \cite{ACG}. To see this, define
\begin{eqnarray}
{\hat{g}} = (k_1 x + k_2)^{-2} g
\end{eqnarray}
and define the co-ordinate $y$,  and the function $Q(y)$, by
\begin{eqnarray}
(n_1 k_2 - n_2 k_1) y &=& {n_1 x + n_2 \over k_1 x + k_2}, 
\nonumber \\
(k_1 x + k_2)^{-4} W(x)^{-1}  &=& {1 \over 4} \ell^2 (n_1 k_2 - k_1 n_2) Q(y)
\end{eqnarray}
where we here assume that\footnote{In the special case for
which  $n_1 k_2 - k_1 n_2=0$, ${\hat{g}}$ corresponds to the product of two
Riemann surfaces.} $n_1 k_2 - k_1 n_2 \neq 0$.
Then
\begin{eqnarray}
{\hat{g}} = y ds^2_{\Sigma} + y Q^{-1} dy^2 + y^{-1} Q (d \psi+ \phi)^2
\end{eqnarray}
where $Q(y)$ is a product of two quadratic polynomials in $y$ and
\begin{eqnarray}
d \phi = {\rm dvol}_{\Sigma}, \qquad
ds^2_{\Sigma} = {1 \over 2} \ell^2 (n_1 k_2 - n_2 k_1) ds^2 (M_2) \ .
\end{eqnarray}
This form of the metric is a special case of the metric appearing in equation (10)
of \cite{ACG}.

It remains to consider the case for which the solution preserves 3/4 of the supersymmetry.
The analysis in this case proceeds using spinorial geometry techniques analogous to
those used to prove that there are no solutions preserving 31 supersymmetries in 
IIB supergravity \cite{preon1}. In particular, by introducing a $Spin(4)$-invariant inner product
on the space of spinors, a 3/4 supersymmetric solution must have spinors orthogonal,
with respect to this inner product,
to a normal spinor. By applying appropriately chosen $Spin(4)$ gauge transformations,
this normal spinor can be reduced to the simple canonical form
as that adapted for the Killing spinor in \cite{DGST11_2}.
In this gauge, a simplified basis for the space of spinors can be chosen. On evaluating the
integrability conditions on this basis, one finds that $F=0$, and the metric must be 
locally isometric to $H^4$.

\section{Supersymmetric Solutions with $SU(2)\times U(1)$ symmetry}

A class of supersymmetric solutions with $SU(2) \times U(1)$ symmetry was
constructed in \cite{sparks3}. In this section, we investigate the relationship
between these solutions, and those found in this work, and in \cite{DGST11_2}.
The solutions in \cite{sparks3} take $\ell=1$, with metric and gauge field strength:
\begin{eqnarray}
\label{mtsol}
ds^2 &=& {r^2-s^2 \over \Omega(r)} dt^2+(r^2-s^2)(\sigma_1^2+\sigma_2^2)
+{4s^2 \Omega(r) \over r^2-s^2} \sigma_3^2
\nonumber \\
F &=& d \bigg( \bigg(P{r^2+s^2 \over r^2-s^2}-Q {2rs \over r^2-s^2}\bigg) \sigma_3 \bigg)
\end{eqnarray}
where 
\begin{eqnarray}
\Omega(r) = (r^2-s^2)^2+(1-4s^2)(r^2+s^2)-2Mr+P^2-Q^2
\end{eqnarray}
and
\begin{eqnarray}
\sigma_1+i\sigma_2 = e^{-i \psi}(d \theta+ i \sin \theta d \phi),
\quad \sigma_3 = d \psi + \cos \theta d \phi \ .
\end{eqnarray} 
The constants $M, s$ are real, and $P$, $Q$ are either both real or both imaginary.
There are two possibilities for supersymmetric solutions. For half-supersymmetric solutions we take
\begin{eqnarray}
\label{halfsusy}
M = Q \sqrt{4s^2-1}, \quad P = -s \sqrt{4s^2-1}
\end{eqnarray}
whereas for quarter-supersymmetric solutions
\begin{eqnarray}
\label{quartersusy}
M=2sQ, \quad P=-{1 \over 2}(4s^2-1) \ .
\end{eqnarray}

\subsection{Half-Supersymmetric $SU(2) \times U(1)$-symmetric Solutions}

For the half-supersymmetric solutions satisfying ({\ref{halfsusy}}), 
there are two cases, according as $s^2<{1 \over 4}$ and $s^2 > {1 \over 4}$.
We begin with the case $s^2<{1 \over 4}$; in this case $P$ and $Q$ are both imaginary.
It is useful to set
\begin{eqnarray}
\label{param}
s = {1 \over 2} \cos \mu, \quad M={k \over \sqrt{2}} \sin \mu,
\quad P=-{i \over 2} \sin \mu \cos \mu, \quad Q=-{i \over \sqrt{2}} k
\end{eqnarray}
for real constants $\mu, k$.
To proceed we introduce new co-ordinates $\sigma$ and $\psi'$ by
\begin{eqnarray}
r=s \bigg({1+\sigma^2 \over 1-\sigma^2}\bigg)
\end{eqnarray}
and
\begin{eqnarray}
\label{shift}
d \psi' = d \psi + f (\sigma)d \sigma
\end{eqnarray}
where
\begin{eqnarray}
\label{ffexp}
f(\sigma) = \bigg({2 \sigma^2 \cos^4 \mu  \over (1-\sigma^2)^2}
+{1 \over 2} \sin^2 \mu \cos^2 \mu (\sigma+\sigma^{-1})^2
- \sqrt{2} k \sin \mu \cos \mu (\sigma^{-2}-\sigma^2)
\nonumber \\
+k^2(\sigma^{-1} - \sigma)^2 \bigg)^{-1}
\times \bigg({2 \sin \mu \cos \mu (\sigma^{-1}+ \sigma) \over (\sigma^2-1)}
+2 \sqrt{2} k \sigma^{-1}\bigg) \ .
\end{eqnarray}
After some manipulation, the resulting metric can be 
written as
\begin{eqnarray}
\label{gsol}
ds^2 = {1 \over 2} \bigg({2 \sqrt{2} \over \sigma^2-1}d \sigma
-{1 \over \sqrt{2}}(\sigma^{-1}+\sigma) {\cal{B}}+(\sigma^{-1}-\sigma)\xi
\bigg)^2 +{\sigma^2 \over (1-\sigma^2)^2} ds^2_{GT}
\end{eqnarray}
where $ds^2_{GT}$ is the metric of the Berger sphere given by
\begin{eqnarray}
ds^2_{GT} = \cos^2 \mu \bigg( \cos^2 \mu (\sigma_3')^2+d \theta^2+\sin^2 \theta d \phi^2 \bigg)
\end{eqnarray}
and
\begin{eqnarray}
{\cal{B}} = \sin \mu \cos \mu \sigma_3', \qquad \xi = k \sigma_3'
\end{eqnarray}
where
\begin{eqnarray}
\sigma_3 = d \psi' + \cos \theta d \phi \ .
\end{eqnarray}
The gauge field strength is also given (now using the conventions of
\cite{DGST11_2}) by{\footnote{In order to match
the $s^2<{1 \over 4}$ solution of \cite{sparks3} to those found in the classification of section 4.2 in \cite{DGST11_2}
we take the gauge field strength here to be real, and
the coefficient multiplying $F$ in the Killing spinor equation is also real.
This is to be compared with the $s^2<{1 \over 4}$ solution of 
\cite{sparks3}, where the coefficient multiplying the gauge field strength
in the Killing spinor equation is imaginary, but the gauge field strength is also imaginary.}
\begin{eqnarray}
F = d \bigg(-{1 \over 4}(\sigma^2+\sigma^{-2}){\cal{B}}+{1 \over 2 \sqrt{2}}
(\sigma^{-2}-\sigma^2) \xi \bigg) \ .
\end{eqnarray}
As expected, it follows that the half-supersymmetric solutions of \cite{sparks3} with $s^2<{1 \over 4}$ correspond to a special case of the half-supersymmetric solutions found previously in 
section 4.2 of \cite{DGST11_2}.
To obtain the $s^2>{1 \over 4}$ solutions in ({\ref{mtsol}}) one makes an analytic continuation of the parameters
({\ref{param}}) of the form
\begin{eqnarray}
\mu \rightarrow i \mu, \quad k \rightarrow ik \ .
\end{eqnarray}
It is clear that when this is done, the gauge field strength becomes imaginary.
However, when this analytic continuation is applied to the solution ({\ref{gsol}}),
because the function $f$ given in ({\ref{ffexp}}) becomes imaginary, it follows that
$\sigma_3'$ becomes complex, because the co-ordinate
$\psi'$ is complex. Furthermore, the term
\begin{eqnarray}
{2 \sqrt{2} \over \sigma^2-1}d \sigma
-{1 \over \sqrt{2}}(\sigma^{-1}+\sigma) {\cal{B}}+(\sigma^{-1}-\sigma)\xi
\end{eqnarray}
which appears in the metric ({\ref{gsol}}) also becomes complex under the analytic
continuation. So the analytically continued solution 
need {\it not}
lie within the class of solutions constructed in section 3 of \cite{DGST11_2}, or in 
the analysis of the previous section, because in these cases, it is assumed that
the metric is real.

\subsection{Quarter-Supersymmetric $SU(2) \times U(1)$-symmetric Solutions}

Next, we consider the quarter-supersymmetric solutions in \cite{sparks3},
satisfying ({\ref{quartersusy}}). In this case, we begin 
by considering the
solution of the previous section ({\ref{metric_14}}),
with $M_2=S^2$,
\begin{eqnarray}
ds^2_{M_2}={1 \over k_1 n_3+n_1 k_3-k_2 n_2}(d \theta^2
+\sin^2 \theta d \phi^2)
\end{eqnarray}
and assume that
\begin{eqnarray}
\label{assum1}
k_1 n_1 >0, \quad k_1 n_3+n_1 k_3-k_2 n_2>0, \quad
k_1 n_2 - k_2 n_1 \neq 0 \ .
\end{eqnarray}
Then, set $\ell=1$, and change co-ordinates from 
$x$ to $r$ where
\begin{eqnarray}
r={s \over (k_1 n_2 - k_2 n_1)}(2 k_1 n_1 x+k_1 n_2 + k_2 n_1)
\end{eqnarray}
with
\begin{eqnarray}
s^2 &=& {(k_1 n_2 - k_2 n_1)^2 \over 8 k_1 n_1(k_1 n_3+n_1 k_3-k_2 n_2)}
\nonumber \\
M^2 &=& {1 \over 32 (n_1 k_1)^3}
(k_1 n_3 + n_1 k_3 - n_2 k_2)^{-3}
(k_1 n_2 - k_2 n_1)^2
\nonumber \\
&\times& (-2 k_1 n_1^2 k_3 - k_1^2 n_2^2
+ n_1^2 k_2^2 +2 k_1^2 n_1 n_3)^2
\nonumber \\
P &=& {1 \over 4 n_1 k_1 (k_1 n_3+n_1 k_3 - n_2 k_2)}
(n_1^2 k_2^2+k_1^2 n_2^2 -2 n_1 n_3 k_1^2-2 k_1 k_3 n_1^2)
\nonumber \\
Q &=& {1 \over 4 n_1 k_1 (k_1 n_3+n_1 k_3 - n_2 k_2)}
(-n_1^2 k_2^2+k_1^2 n_2^2 -2 n_1 n_3 k_1^2+2 k_1 k_3 n_1^2) \ .
\end{eqnarray}
After some manipulation, one recovers the quarter supersymmetric solutions in \cite{sparks3},
satisfying ({\ref{quartersusy}}); it is straightforward to
see that the gauge field strengths also match.

Hence, although the conditions obtained in the
previous section by considering the integrability of the Killing spinor equations are necessary for the enhancement of supersymmetry, they are not sufficient. In particular,
an explicit construction of the Killing spinors will
produce additional conditions on the constants $n_i, k_i$
which are sufficient to ensure supersymmetry enhancement. 

We shall conclude by outlining a more geometric approach
where the difference between  the necessary and sufficient conditions becomes  apparent. This approach follows \cite{DT13}, where it has been used to characterise four dimensional Riemannian manifolds  conformally equivalent to hyper-Kahler.
 
Equation (\ref{kse}) defines a connection ${\bf D}$ on a rank four complex vector bundle over $M_4$, and Killing spinors are in a 1-1 correspondence with the parallel sections of this connection. The necessary 
conditions for the enhanced supersymmetry analysed in Section 
\ref {section5} arise from commuting the covariant derivatives, and thus computing the curvature ${\bf R}=[{\bf D}, {\bf D}]$ of ${\bf D}$. This curvature, which is
essentially given by (\ref{intg}), can be thought of as a four by four matrix, and the necessary conditions which we have computed are equivalent to the statement that this matrix has two--dimensional kernel. Thus there are two linearly independent solutions to 
\begin{equation}
\label{curvature_of_D}
{\bf R}\epsilon=0.
\end{equation}
To satisfy the sufficient conditions for the SUSY enhancement we need to make sure
that all differential consequences of the formula 
(\ref{curvature_of_D})
hold which, by the
Frobenius theorem, then guarantees that the kernel of ${\bf R}$ is parallel w.r.t the connection ${\bf D}$. 
We differentiate (\ref{curvature_of_D}) and use the parallel property of
sections
${\bf D}\epsilon=0$ to produce a sequence of matrix algebraic conditions
\[
{\bf R}\epsilon=0, \quad  ({\bf D}{\bf R})\epsilon=0, \quad
({\bf D}^2{\bf R})\epsilon=0,\quad \dots\;\;.
\]
We stop the process once the differentiation does not produce new equations. For enhanced supersymmetry 
the rank of the extended matrix $\{ {\bf R}, {\bf D}{\bf R}, \dots\}$
should be two. Therefore, as the rank of ${\bf R}$ is  two when the necessary conditions hold, it is sufficient to demand that
the condition $({\bf D}{\bf R})\epsilon=0$ holds identically.
Thus the vanishing of all 3 by 3 minors of the matrix 
$\{ {\bf R}, {\bf D}{\bf R} \}$ will give the sufficiency conditions for the existence of two Killing spinors.

\acknowledgments

 MD and JG
thank the American University of Beirut
for hospitality when some of this work was carried over.
MD is grateful to Vestislav Apostolov and Chris Pope for useful correspondence,
and to Ian Anderson and Charles Torre for introducing him to their MAPLE
DifferentialGeometry package used to verify the calculations in Section 3.
JG is supported by the STFC grant ST/I004874/1.


\begin{thebibliography}{99}


\bibitem{AG97} V. Apostolov and P. Gauduchon,
\textit{The Riemannian Goldberg-Sachs theorem,  Internat. J. Math.}  {\bf 8} (1997) 
421.

\bibitem{ACG}  V. Apostolov, D. Calderbank and P. Gauduchon, 
\textit{Ambikaehler geometry, ambitoric surfaces and Einstein 4-orbifolds;} [arXiv:1010.0992 [math.DG]].

\bibitem{BF} C. Boyer and D. Finley,
\textit{Killing vectors in self-dual Euclidean Einstein spaces, J. Math. Phys.} {\bf 23} (1982) 1126. 


\bibitem{Der83} A. Derdzi\'nski,  
\textit{Self-dual Kahler manifolds and Einstein manifolds of dimension four,  Compositio Math. }  {\bf 49} (1983) 405. 

\bibitem{Dbook} Dunajski, M.  (2009)
{\it Solitons, Instantons and Twistors}.
Oxford Graduate Texts in Mathematics {\bf 19}, Oxford University Press.

%\bibitem{DT10} Dunajski, M. and Tod, K. P. (2010) 
%Four Dimensional Metrics Conformal to K\"ahler, Math. Proc. Camb. Phil. Soc. {\bf 148}, 485-503.

\bibitem{DH07} M. Dunajski and S. A. Hartnoll, 
\textit{Einstein-Maxwell gravitational instantons and five dimensional solitonic strings, Class. Quantum. Grav.} {\bf{24}} (2007) 1841; [hep-th/0610261].

\bibitem{DGST11} M. Dunajski, J. B. Gutowski, W. A.  Sabra and K. P. Tod,  
\textit{Cosmological Einstein-Maxwell Instantons and Euclidean
Supersymmetry: Anti-Self-Dual Solutions, 
Class. Quant. Grav. } {\bf 28.} (2011) 025007; [arXiv:1006.5149 [hep-th]].

\bibitem{DGST11_2}
M. Dunajski, J. B. Gutowski, W. A. Sabra and K. P. Tod,
\textit{Cosmological Einstein-Maxwell instantons and Euclidean supersymmetry: beyond self-duality, 
JHEP} {\bf{03}} (2011) 131; [arXiv:1012.1326 [hep-th]].

\bibitem{DT13} M. Dunajski and K. P. Tod,  
\textit{Self--Dual Conformal Gravity}; 
[arXiv:1304.7772 [hep-th]].
  
 \bibitem{d11integ}
 J. M. Figueroa-O'Farrill  and G Papadopoulos,
  \textit{Maximally supersymmetric solutions of ten-dimensional and eleven-dimensional supergravities,
  JHEP} {\bf 03} (2003) 048; [hep-th/0211089].


  
  \bibitem{d11class1}
  J. P. Gauntlett  and S. Pakis,
  \textit{The Geometry of D = 11 killing spinors,
  JHEP} {\bf 04} (2003) 039; [hep-th/0212008].
  
  \bibitem{d11class2}
  J. P. Gauntlett, J. B. Gutowski and S. Pakis,
  \textit{The Geometry of D = 11 null Killing spinors,
  JHEP} {\bf 12} (2003)  049; [hep-th/0311112].


\bibitem{gary_rych} G. W. Gibbons and P. Rychenkova, 
\textit{Single-sided domain walls in M-theory, J. Geom. Phys. } {\bf 32} (2000) 311; [hep-th/9811045].

\bibitem{preon1}
U. Gran, J. Gutowski, G. Papadopoulos and D. Roest,
\textit{N=31 is not IIB, JHEP} {\bf 02} (2007) 044; [hep-th/0606049].

\bibitem{systematics}
 U. Gran, G. Papadopoulos and D. Roest 
  \textit{Systematics of M-theory spinorial geometry,
  Class.\ Quant.\ Grav.\ } {\bf 22} (2005)  2701; [hep-th/0503046].
 


\bibitem{GT10}
J. B. Gutowski and W. A.  Sabra,
\textit{Gravitational Instantons and Euclidean Supersymmetry.
Phys. Lett.} {\bf B693} (2010) 498; [arXiv:1007.2421 [hep-th]].


\bibitem{lukier} J. Lukierski and  W. J. Zakrzewski,
\textit{Euclidean supersymmetrization of instantons and self-dual monopoles, 
Phys. Lett. } {\bf B189} (1987) 99.

\bibitem{sparks1} D. Martelli and J. Sparks,
\textit{The nuts and bolts of supersymmetric gauge theories on biaxially squashed 
three-spheres, Nucl.\ Phys.\ } {\bf B866} (2013) 72; [arXiv:1111.6930 [hep-th]].

\bibitem{sparks2} D. Martelli, A. Passias and J. Sparks, 
\textit{The gravity dual of supersymmetric gauge theories on a squashed three-sphere,
Nucl.\ Phys.\ } {\bf B864} (2012) 840; [arXiv:1110.6400 [hep-th]]. 

\bibitem{sparks3} D. Martelli, A. Passias and J. Sparks,
\textit{The supersymmetric NUTs and bolts of holography;} [arXiv:1212.4618 [hep-th]].

\bibitem{MSW} L. Martina, M. B.  Sheftel and P. Winternitz,
\textit{Group foliation and non-invariant solutions of the heavenly equation, J. Phys. }
{\bf A34} (2001) 9243; [math-ph/0108004].


\bibitem{Pope78} C. N. Pope, 
\textit{ Eigenfunctions and ${\rm Spin}^{c}$ structures in ${\mathbb{C}}P^2$, 
Phys. Lett. } {\bf B 97} (1980) 417. 

\bibitem{Pope_lift} C. N. Pope,
\textit{Consistency of truncations in Kaluza-Klein,
The Santa Fe Meeting},
editors T. Goldman  and M. M. Nieto, World Scientific (1984)

\bibitem{Pope3} C. N. Pope, A. Sadrzadeh and  S. R. Scuro,
\textit{Timelike Hopf Duality and Type $IIA^*$ String Solutions.
Class. Quant. Grav. } {\bf 17} (2000) 623; [hep-th/9905161].

\bibitem{T97v1} K. P. Tod, 
\textit{The ${\rm SU}(\infty)$-Toda field equation and special four-dimensional
metrics, Geometry and physics} Aarhus (1995).

\bibitem{T97v2} K. P. Tod,
\textit{Lecture Notes in Pure and Appl. Math.}, Dekker, New York, (1997) 184.

\bibitem{T95}  K. P. Tod, 
\textit{Scalar-flat K\"ahler and hyper-K\"ahler
metrics from Painlev\'e III, Class. Quant. Grav } {\bf 12} (1995) 1535.


\end{thebibliography}
\end{document}